\documentclass[conference]{IEEEtran}
\usepackage{graphicx}
\usepackage{amsmath}
\usepackage{algpseudocode}
\usepackage{algorithm}
\usepackage{paralist}

\newcommand{\rom}[1]{\uppercase\expandafter{\romannumeral #1\relax}}

\makeatletter
\def\ps@headings{%
\def\@oddhead{\mbox{}\scriptsize\rightmark \hfil \thepage}%
\def\@evenhead{\scriptsize\thepage \hfil \leftmark\mbox{}}%
\def\@oddfoot{}%
\def\@evenfoot{}}
\makeatother
\begin{document}
%
\title{Content Based Traffic Engineering in Software Defined Information Centric Networks}


\author{\IEEEauthorblockN{\hspace{2.5cm} Abhishek Chanda\IEEEauthorrefmark{1}\vspace{0.25cm}}
\IEEEauthorblockA{\IEEEauthorrefmark{1}WINLAB\\ Rutgers University\\
New Brunswick, NJ 08901\\
\{achanda, ray\}@winlab.rutgers.edu}
\and
\IEEEauthorblockN{Cedric Westphal\IEEEauthorrefmark{2}\IEEEauthorrefmark{3}\vspace{0.25cm}}
\IEEEauthorblockA{\IEEEauthorrefmark{2}Innovation Center\\ Huawei\\
Santa Clara, CA 95050\\
cedric.westphal@huawei.com}
\and
\IEEEauthorblockN{Dipankar Raychaudhuri\IEEEauthorrefmark{1} \hspace{2.5cm}\vspace{0.25cm}}
\IEEEauthorblockA{\IEEEauthorrefmark{3}Dept of Computer Engineering\\ University of California, Santa Cruz\\
Santa Cruz, CA 95064\\
cedric@soe.ucsc.edu}
}
\maketitle

\begin{abstract}
This paper describes a content centric network
architecture which uses software defined networking principles
to implement efficient metadata driven services by extracting
content metadata at the network layer. The ability to access
content metadata transparently enables a number of new services
in the network. Specific examples discussed here include: a
metadata driven traffic engineering scheme which uses prior
knowledge of content length to optimize content delivery, a
metadata driven content firewall which is more resilient than
traditional firewalls and differentiated treatment of content based
on the type of content being accessed. A detailed outline of an
implementation of the proposed architecture is presented along
with some basic evaluation.
\end{abstract}
\IEEEpeerreviewmaketitle

\section{Introduction}
\label{sec:intro}

Information Centric Network (ICN) architectures aim to
alleviate the problems associated with traditional networks by
operating on content at different levels. An ICN uses content
names to provide network services like routing and content
delivery. To facilitate this, a typical ICN will need to have a
content management layer to handle routing based on content
names. In an ICN, some nodes are assumed to have different
levels of temporary storage. It is typical for an ICN node to
provide caching, where these nodes can store content indexed
by content name. This caching is transparent to the end-users,
who are agnostic to the content location. This often reduces
the access latency for content delivery and increases network
efficiency. These design patterns require that other network
services like traffic engineering, load balancing etc. should be
done with content names and not with routable addresses as
well. This current work is inspired by the observation that
in an ICN, various information about content can be derived
by observing in-network content flows, or content state in the
cache (or by using \emph{deep packet inspection} mechanism at the
switches).

Operating on content introduces new opportunities, which
we have not seen explored yet by the networking community.
Unlike an IP \emph{flow} in a traditional network which does not
have an explicit end marker\footnote{The flow can time-out, which is an implicit end, but requires the proper time-out value.}, content has explicit beginning
and end semantics which makes it easy to provide the proper amount of resource for the flow, and track how much data has
passed through a device. The ability to detect these explicit
events helps us set up a network such that it will only let the
desired content pass and the resource will automatically be
de-allocated once the content flow has ended.



In this paper, we propose a scalable architecture for exploiting
the explicitly finite nature of content semantics. We present
a mechanism to observe and extract content metadata at the
network layer, and use this to optimize the network behavior.
In particular, we suggest three illustrative applications of this,
but we are confident that this expressiveness of the ICN
architecture would support many more:
\begin{inparaenum}
\item Supporting traffic engineering in an ICN
\item Supporting a content firewall service
\item Performing network-wide cache management based on a
function of size and popularity.
\end{inparaenum}
We show that extracting
content metadata \emph{for free} (as a by-product of the ICN
paradigm) enables us to support metadata driven services in
a network. We subscribe to the emerging software defined
networking (SDN) \cite{nick08} philosophy of forwarding and control
plane separation and demonstrate a possible implementation
of the proposed architecture on a standard SDN platform.
Essentially, we show how an existing SDN control plane can
be augmented to include a content management layer which
supports TE and firewalling. Our proposal does not need any
application layer involvement.

The rest of the paper is organized as follows: section \ref{sec:related} presents some related work, section \ref{sec:network} presents the architecture for our proposed scheme and describes the set of services the given architecture can enable, section\ref{sec:implementation} describes modifications necessary to OpenFlow to support these services, section \ref{sec:optimization} shows an optimization problem that can be used here, section \ref{sec:eval} shows results from a simple evaluation, finally section \ref{sec:conclusion} concludes the paper.

\begin{figure}
\centering
\vspace{-30pt}
\includegraphics[width=90mm,trim=0 200 0 0]{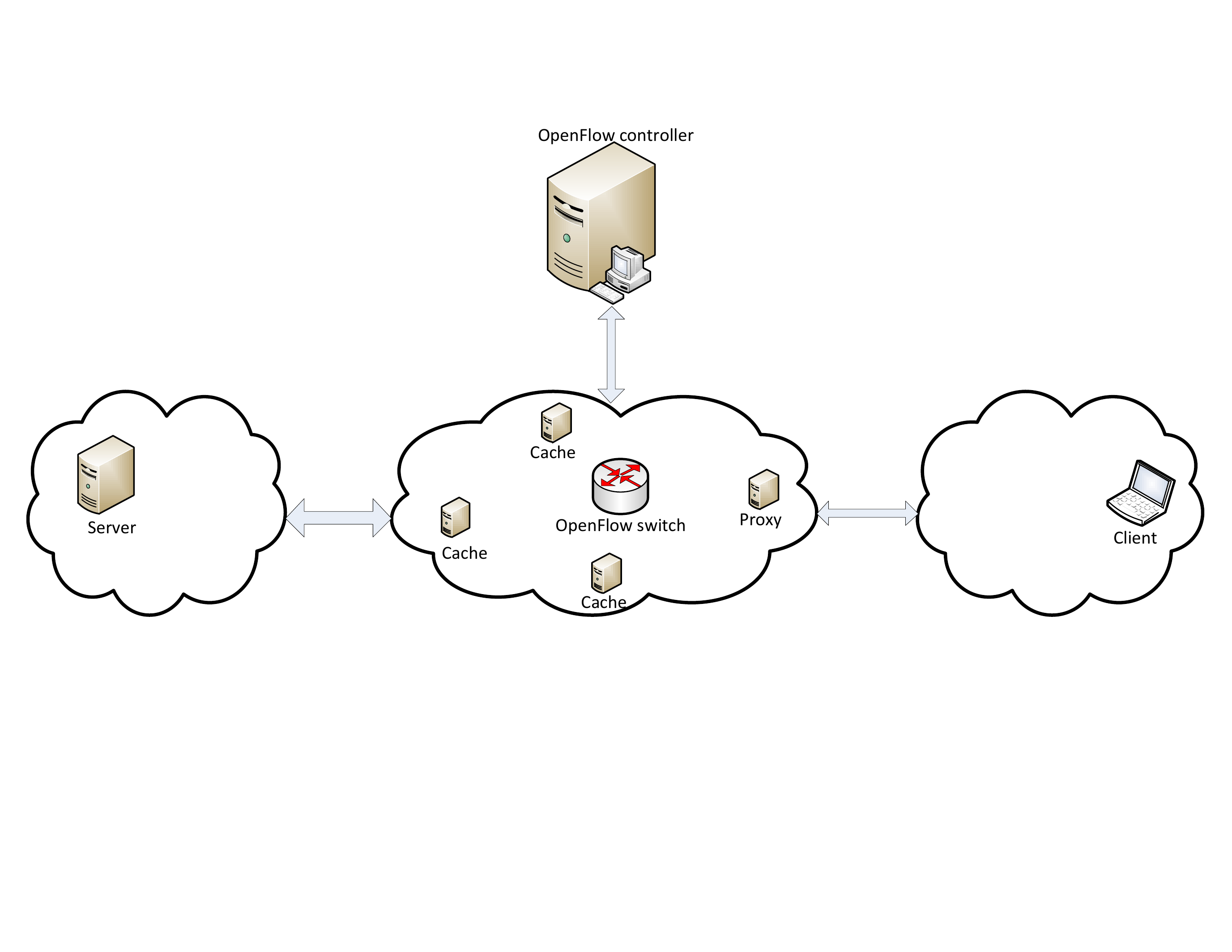}
\caption{Network model}
\label{fig:network}
\end{figure}

\section{Related work}
\label{sec:related}
We present some existing work that are relevant to our
documents. The problem of traffic engineering has been
studied extensively since the early days of the Internet and
telecom networks. RFC 3272 provides an overview of TE in
the Internet \cite{Awduche02} and also defines traffic engineering in this
context. Reference \cite{Fortz02} is another work which describes TE in
an IP network and shows that traditional shortest path routing protocols can be used for TE in IP networks by instrumenting
link weights based on traffic.

More recently, with the advent of cloud computing, a lot
of data intensive applications have moved to data centers.
Thus, traffic engineering has found applications in data center
networking with its own set of challenges. A number of
authors have identified the merits of traffic engineering of
\emph{elephant flows} \cite{Curtis11}, \cite{Fares10} -– flows that carry a large amount of data -– in a data center network. Curtis et. al. proposed Mahout \cite{Curtis11}
which aims to identify elephant flows by installing a shim layer
on end hosts. The layer monitors TCP send buffer sizes on the
hosts and reports them back back to an OpenFlow controller.
The controller can aggregate all statistics collected from a
path to identify elephant flows on that path. Those flows can
then be re-allocated according to some optimization criteria.
One major issue with this architecture is that it requires
modification of end hosts, which is not always desirable.
In this work, we keep track of content in a similar manner
as Mahout keeps track of elephant flows in the data center.
\cite{Fares10} proposed Hedera on similar grounds which works by
instrumenting end switches in a data center network. None
of these works focus on ICN and thus fail to take advantage
of the extra information associated with knowledge of content
properties above the transport layer. For instance, identifying
elephant content in an ICN does not require any end point
support.

Very recently, some authors have started to investigate TE
in an ICN. \cite{Frank12} proposed Content
Aware Traffic Engineering (CaTE)which works by exploiting
path diversity content information in a ISP network. Xie et.
al. proposed a method to decouple traffic engineering and
collaborative caching in a network \cite{Xie11}. Chanda et. al. proposed
the inclusion of non-forwarding elements which are essential
in an ICN under a centralized control layer \cite{Chanda12}. Our present
work is a natural extension of this work.

SDN is a new trend in networking that advocates the
separation of the control plane and the forwarding plane in
a network. This makes the control plane programmable and
enables writing a network operating system that can manage
a set of networking resources. The most prominent SDN
technology is OpenFlow \cite{nick08}. While we find this approach very
useful for our evaluation and implementation, we notice that
per flow granularity is not enough to operate on content. Thus,
we propose a few modifications to OpenFlow to enable it to support content operations seamlessly, which we discuss later.

\section{Network model and architecture}
\label{sec:network}
\subsection{Overview}
We take the point of view of a network operator, and assume
that content is requested by an end-user from a server, both of
which can be outside of the network. We assume this network
operates with a control plane which manages content. Namely,
when a content request arrives in the network, the control
plane locates the proper copy of the content (internally in
a cache, or externally from its origin server); when content
objects arrive in the network, the control plane has the ability
to route the content, and to also fork the content flow towards
a cache (on the path or off the path). The control plane can
leverage ICN semantics (say, CCN \emph{interest} and \emph{data} packets)
to identify content, or can be built upon existing networks,
for instance using the SDN concepts as in \cite{Chanda12}. Our proposal
works in either context, but is described as built upon \cite{Chanda12} so
that we can use legacy clients and servers to integrate with
the caching network.

Figure 1 shows the placement of such a network. It is
an OpenFlow network which is managed by a controller
(without loss of generality, we can assume only one controller
for the network). The controller runs a content management
layer (within the control plane) that manages content names,
translates them to routable addresses and also manages caching
policies and traffic engineering. The control plane translates
the information from the content layer to flow rules which are
pushed down to switches. Some of the switches in the network
have the extra ability to parse content metadata and pass on to
the content management layer in the controller.We assume that
the service provider network connects to the caching network
using one or more designated ingress switches.

Figure 2 shows the interaction between the augmented
control plane and the forwarding plane. The content management
layer has a number of modules for each task as
shown in the figure. The enhanced forwarding plane sends
back content metadata to the controller, which is used to
make forwarding decisions. The control plane pushes back
flows to the forwarding plane. Thus the whole systems forms
a closed feedback loop. One key element of this architecture is
a key-value store in the content metadata manager which maps
the (globally unique) content name to some network-extracted
metadata. In particular, we will demonstrate the benefit of only
keeping in the key-value store the content size.

The network also includes a number of cache and proxy
nodes which can talk to the OpenFlow controller and announce
their capability. Thus, the controller can decide to cache a
content in a selected location (based on some optimization
criteria). The proxy nodes can be used to transparently demultiplex
TCP connections between caches. However, we need
some extra functionality which are described below.
\begin{figure}
\centering
\vspace{-10mm}
\includegraphics[scale=0.35]{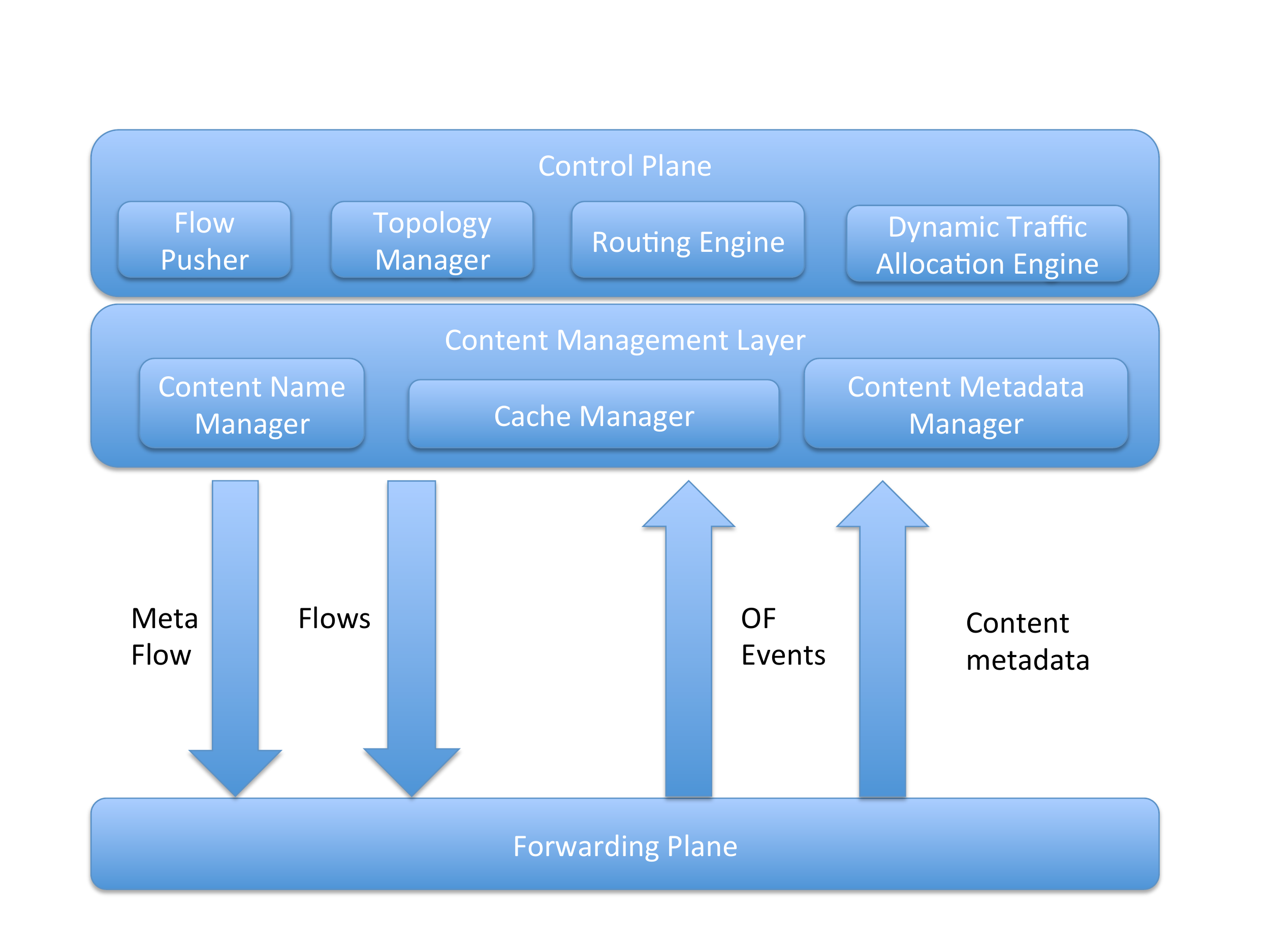}
\caption{Metadata feedback loop}
\label{fig:feedback}
\end{figure}
\subsection{Content metadata extraction}
Since we intend to implement TE and firewalling by using
content metadata, we will need a mechanism to extract that information. We can define two levels of abstraction in this context. The first one is strictly at the network layer, and takes
advantage of the ICN semantics. The second goes into the
application layer.
\begin{itemize}
\item \emph{Network-layer mechanism to extract content length:} Since
in an ICN, content is uniquely identifiable, the controller
can recognize requests for new content (i.e. content for
which the controller holds no metadata in the key-value
store). For new content, the controller can set up a
counter at a switch (say, the ingress switch) to count the
content flow size\footnote{In this case, the flow size will include some header overhead that the controller can subtract later on.}. The controller can also request that the
flow be cached, and obtain the full object size from
the cache memory footprint. When the content travels
through the network later, a look-up to the key-value
store allows to allocate resource accordingly. Further, the
content that is observed for the first time can still be
classified into \emph{elephant} and \emph{mice} flows on the fly based
on some threshold and allocated accordingly to optimize
some constraint.
\item \emph{Application-layer mechanism to read HTTP headers in
the ingress switch:} In the previous case, content size is
not available for content which is observed for the first
time. However, this can be extracted by parsing the HTTP
headers. This will allow the controller to do elephant flow
detection and take appropriate actions early. Though it is
difficult to implement, an advantage of this method is that
it will allow us to do TE and firewalling from the first
content occurrence.
\end{itemize}

\subsection{Controller driven coordination}
Once the network elements have the ability to extract content
metadata, they will need to announce their ability to the
controller. We propose that this should be done in-band using
the OpenFlow protocol since it supports device registration and
announcement features. This essentially involves the following
steps
\begin{itemize}
\item Asynchronous presence announcement, where a device announces it's presence by sending a hello message to the assigned controller.
\item Synchronous feature query, where the controller acknowledges the device's announcement and asks it to advertise its features.
\item Synchronous feature reply, when the device replies with a list of features.
\end{itemize}
After these steps, the controller has established sessions to all devices and knows their capabilities, it can then program devices as necessary.

Given the setup described, the controller can have extra information about content in a network. Also, the SDN paradigm allows the controller to have a global view of the network. Thus, the platform can support implementation of a number of services. Here, we will discuss a few services as example.
\begin{itemize}
\item \emph{Metadata driven traffic engineering}
Amongst various content metadata parameters, since the controller knows the content length, it can solve an optimization problem under a set of constraints to derive paths on which the content should be forwarded. Modern networks often have path diversity between two given devices. This property can be exploited to do traffic engineering. This approach of doing traffic engineering is efficient and scalable since it does not require the service providers to transfer content metadata separately, which saves network bandwidth at both ends.

\item \emph{Differentiated content handling}
The DPI driven mechanism
described earlier enables us to have rich content
metadata. Thus, assuming such a content metadata extraction
service, the content management layer in the
controller can take forwarding decisions based on the
MIME type of the content. A network administrator can
describe a set of policies based on content types. Let us
take delay bound for example. If the MIME type is that of
a real time streaming content (a video clip), it can select
a path which meets the delivery constraints (the delay
bound which has been set). If none of the paths can satisfy
the bound, the path with the lowest excess delay will be
selected. This approach can be combined with the traffic
engineering described above to handle multiple streaming
content on a switch by selecting different paths for each.

\item \emph{Metadata driven content firewall}
We can envision a
metadata driven content firewall in the network. When a
piece of content starts to enter the network, the controller
knows how long is it. Thus, it should be possible to
terminate all flows handling the content after the given
amount of data has been exchanged. This mechanism
acts like a firewall in the sense that it opens up the
network only to transmit the required amount of data.
We argue that this mechanism provides stronger security
than traditional firewalls. With a traditional firewall, the
network administrator can block a set of addresses (or
some other parameters). But the attacker can always spoof IP addresses and bypass the firewall. However, with this
proposed mechanism, the network will not let content
from spoofed IP addresses pass through since it knows
that the content has already been transmitted.
\item \emph{Metadata driven cache management}
As object size varies
in the cache, the cache policy needs to know not only the
popularity of the content and its frequency of access, but
also its size, to determine the best ”bang for the buck” in
keeping the content. Having access to the content requests
and the content size at the controller provide both.
\end{itemize}
Given the architecture, we can show the end to end flow
of content in the network, in a typical scenario. Consider a
system as shown in figure 3. In this example we will assume
that the objective is to optimize link bandwidth utilization by
load balancing incoming content across redundant paths. We
show both the schemes described in section III in the diagram.
However, in a real implementation, it is sufficient to have one.
Here are the steps the system will take, the initial few steps
are very similar to that described in [8]. Note that this can
be subdivided into three distinct steps: the first step is the
setup phase where all devices connect to the controller and
announce their capabilities, the second phase in the metadata
gathering phase where network elements report back content
metadata, this is used in the third phase for TE.
\begin{itemize}
\item All elements boot up and connect to the controller.
\item Elements announce their capabilities to the controller. At
this point, the controller has a map of the whole network
and it also knows which nodes can extract metadata and
cache content.
\item Controller writes the special flow in all ingress switches,
configuring them to extract content metadata. It also
writes a ”flow” to the cache asking it to report back
content metadata.
\item The client tries to setup a TCP connection to a server in
the content provider network.
\item An OpenFlow switch in the content network forwards the
packets to the controller; which writes flows to redirect
all packets from client to a proxy. At this stage, the client
is transparently connected to the proxy.
\item The client sends a GET request for a piece of content.
Proxy parses the request and queries controller to see if
that content is cached in the network.
\item The first request for a content will be a cache miss
since it is not already cached. Thus the controller returns
nothing, the proxy forwards the request to the server in
the provider network.
\item The server sends back the content which reaches the
ingress switch. The switch asks the controller where the
content should be cached. This marks the explicit start of
content.
\item A special flow is pushed to each of the switches in the
path and the content is cached. At this point, the controller
knows where a content is cached.
\item The ingress switch and the cached both returns content
metadata. Now, the controller can map this metadata to a content name.
\item If another consumer requests for the same content, the
controller looks up it’s cache dictionary by content name
and the proxy redirects the request to the cache. Simultaneously,
the controller uses the TE module to compute a
path on which the content should be pushed to improve
overall bandwidth utilization in the network. It writes
flows to all switches to forward content.
\end{itemize}
In this case, the following algorithm is used in the controller:
\begin{algorithm}
\begin{algorithmic}
    \State tempDictionary = null
    \State P = get all routes from ingress switch to selected cache
    \For{$p~in~P$}
        \State $tempCost = 0$
        \For{$e~in~p$}
            \State $tempCost = tempCost + \frac{b_e + F}{c_e}$
        \EndFor
        \State insert $\langle p,tempCost\rangle$ in tempDictionary
    \EndFor
    \State return the path corresponding to the minimum cost in tempDictionary
\end{algorithmic}
\caption{Example of a typical path selection algorithm}
\end{algorithm}

As mentioned before, the actual optimization algorithm to
be used depends on the problem definition and can be changed.

\section{Implementation discussions}
\label{sec:implementation}
As it is not an ICN architecture, the standard OpenFlow does
not support all the functionality outlined in the earlier section.
Here, we describe necessary modifications to the OpenFlow
protocol to support the proposed mechanisms, since we view
this as the fastest path to implementation. In this paper, we
will focus only on content sent over HTTP since it forms
the majority of Internet traffic. Other types of content can
be addressed by extending the proposed mechanism in future
work.

From a top level, network elements need to announce their
capability of parsing (and caching) content metadata to the
controller. Then the controller should be able to write flows
which will configure them to parse and send back metadata to
the controller. Necessary modifications are described below:
\subsection{Switch-controller handshake}
During the handshake phase, the switch needs to announce its capability to parse content metadata. The controller can maintain a table of all switches that has advertised this capability.The switch announces its capabilities in a $OFPT\_FEATURES\_REPLY$ message. So, we will need to add extra fields to the $ofp\_capabilities$ structure indicating capabilities to extract content metadata, cache content and proxy content.
\subsection{Augmented flowmod}
Once the controller connects to all the elements in the network, it will know which elements can extract metadata. The control plane will need to configure those elements by writing flowmods, asking them to parse content metadata. Thus, we will need an additional action on top of OpenFlow. We call it $EXTRACT\_METADATA$. A flowmod with this action will look like this
\begin{verbatim}
if ; actions=EXTRACT_METADATA,NORMAL
\end{verbatim}
which essentially means, the switch will extract metadata from HTTP metadata, put it in a $PACKET\_IN$ message and send back to the controller. Later, the switch will do a normal forwarding action on the packet.

We also introduce a new type of flowmod to OpenFlow. This format provides the ability to write flowmods which have an expiry condition:
\begin{verbatim}
if <conditions>; actions=<set of actions>
;until=<set of conditions>
\end{verbatim}
Now, since the controller knows the length of a given content, it can use the per flow byte counter to set a condition for the
\emph{until} clause.
\subsection{Switch to controller message}
We are mostly interested in the content length which is encoded in HTTP headers (note that it is easy to extend this mechanism to extract other content metadata like mime type etc). Once a switch is configured to parse content, when it sees a HTTP packet, it will read the \verb Content-length  header and construct a tuple of the form $\langle content\:name,content\:size,src\:ip,src\:port,dest\:ip,dest\:port\rangle$. This will be encapsulated in a $PACKET\_IN$ and sent back to the controller.
\subsection{Content management layer implementation}
Most OpenFlow controller allow a module system and a mechanism for modules to listen on $PACKET\_IN$ messages. Thus, the control management layer can be implemented as a module on a controller. It will subscribe to $PACKET\_IN$ messages. When it gets a packet, it will extract the information and discard the packet. This architecture allows the controller side to have multiple content management layers chained together.

The interaction between the switch and the controller is shown in the figure~\ref{fig:network}. Modified OpenFlow messages are marked with a star.


\section{Optimization problem formulation}
\label{sec:optimization}
To demonstrate the power of our approach, we focus now
on network traffic optimization. The goal here is to optimize
some metric of the network using content metadata that is
gathered through content centric hooks in the network and
is available to the controller. Let us split the problem into
two parts. The first sub-problem is storing the content in a
cache: when the controller selects a cache, it will have to
select a path to the cache, and will have to pick a cache that
is not under stress already (for instance, from many write
IOs form receiving other contents). Assuming the network
will have a number of alternate paths between the ingress
switch and the selected cache, this is an opportunity to use
path diversity to optimize link utilization. We also assume
that content metadata is available right after content enters
the network (i.e. the ingress switch does DPI). Thus, here our
objective is to minimize the maximum link utilization. This
means that, we need to solve the following problem,
\begin{equation*}
\begin{aligned}
& \underset{}{\text{min max}}
& \sum_{p \in P} \sum_{e \in p} \frac{b_e + F}{c_e}\\
& \text{subject to}
& b_e \le c_e
\end{aligned}
\end{equation*}

The second sub problem is content retrieval. The goal here is to minimize the delay an user will see when it requests content. This can be formulated as
\begin{equation*}
\begin{aligned}
& \underset{e \in E}{\text{min}}
& \frac{F}{r_e}
\end{aligned}
\end{equation*}

The following table summarizes the notations used:
\begin{center}
  \begin{tabular}{| c | c |}
    \hline
    $b_e$ & Background traffic on link $e$ \\ \hline
    $c_e$ & Capacity of link $e$ \\ \hline
    $r_e$ & Rate of link $e$ \\ \hline
    $F$ & Size of the content \\ \hline
    $P$ & Set of all paths between a source and a destination\\ \hline
    $E$ & Set of all links\\
    \hline
  \end{tabular}
\end{center}
Another interesting optimization problem that can be considered
here is that of disk IO optimization. Given a number of
caches in the network, each having a known amount of load
at a given time, we might want to optimize on the disk writes
over all of them and formulate the problem on that metric.
Note that the actual optimization constraint to be used can
vary on application requirements and is user programmable.
Typically, these constraints will be programmed in the content
management layer of the controller.

\section{Evaluation}
\label{sec:eval}
\begin{figure*}
\centering
\includegraphics[width=170mm,height=70mm]{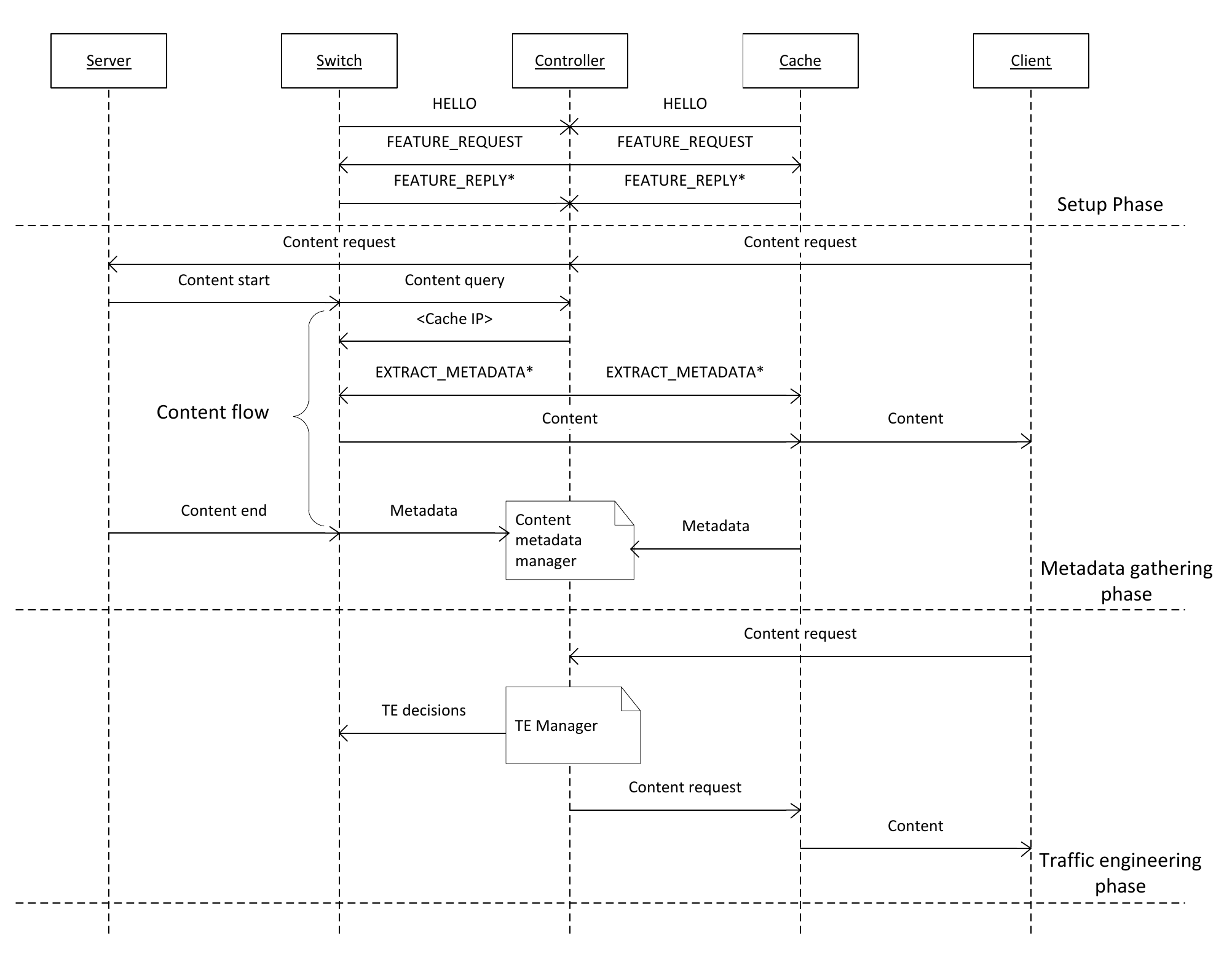}
\caption{End to end flow in the system}
\label{fig:network}
\end{figure*}
In this section, we will demonstrate that the prior knowledge of content size can be used to decrease backlog in a link, which in turn results in better utilization. The setup is the following, we have two parallel links between a source and a destination. Each link has a capacity of $1kbps$. Thus the total capacity of the system is $2kbps$ and the aggregated input should be below the link capacity on average (otherwise the queue will go unstable). According to existing literature, we assume that content size is Pareto distributed \cite{Arlitt97}. Given a value of $\alpha$ we will calculate the value of the shape parameter using the relation $b \le 2\frac{\alpha - 1}{\alpha}$ so that the mean of the distribution is always below $2$. We assume deterministic arrival time of content, once every second from $t=1$ to $t=10000$. Traffic is allocated to each based on one of the following policies:
\begin{itemize}
\item \emph{Policy 1} assumes that the content size is not known prior to allocating links. Thus, at any time instant, if both links are at full capacity, we will pick any one randomly. Otherwise, we pick the empty one.
\item \emph{Policy 2} assumes that we know the content size priori. At any time instant, we will pick the link with minimum backlog.
\end{itemize}
We study the variation of the average backlog in each case with increase in $\alpha$ from $1.1$ to $2.5$. For each value of $\alpha$, we plot on Figure 4 the
difference in \% between the total backlog in the system under
both policies. The size-aware policy 2 always reduces the
amount of data waiting to be transmitted, and thus the delay in
the system. For low loads, there is no need for optimization;
for very high load, the links will always be highly backlogged
and both policies are throughput optimal. We want to operate
in a region where the link utilization is close to 1. Policy 2
shows significant improvements in such a case.
\begin{figure}
\centering
\includegraphics[scale=0.5]{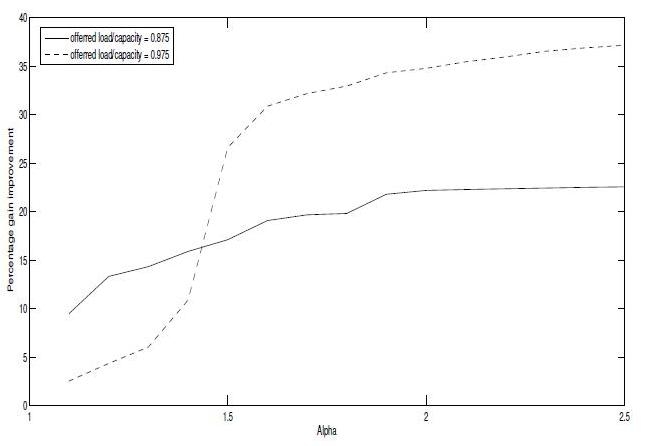}
\caption{Variation of link backlog with increase in $\alpha$: for the load/capacity
of $0.95$, the gain is up to $40\%$ and $26\%$ on average}
\label{fig:result}
\end{figure}

\section{Conclusion}
\label{sec:conclusion}
We have proposed an architecture that aims to support metadata
driven services in an ICN resulting in better utilization of network resources. Here we have instantiated a few specific
services like metadata driven traffic engineering as example.
A natural next step is to implement this proposed scheme at
a larger scale and gather some performance metrics. Another
open question here is a choice of the optimization criteria for a
given network. This can vary widely depending on a network
operator’s constraints, a caching network operator might want
to optimize disk writes while another operator might want to
optimize link bandwidth usage in a data center. We would
like to keep this externally configurable since the architecture
is independent of the underlying optimization problem.

\bibstyle{ieeetr}
\bibliography{bare_conf}

\end{document}